\documentclass[prl,twocolumn,showpacs,preprintnumbers,amsmath,amssymb]{revtex4}

\usepackage{graphicx}
\usepackage{bm}

\begin{document}


\title{Unifying explanation for carrier relaxation anomaly in gapped systems}
\author{Shota Ono}
\email{shota-o@eng.hokudai.ac.jp}
\author{Hiroyuki Shima}%
\author{Yasunori Toda}%
\affiliation{%
Division of Applied Physics, Faculty of Engineering,
Hokkaido University, Sapporo, Hokkaido 060-8628, Japan
}%

\date{\today}

\begin{abstract}
We develop a theory to describe energy relaxation of photo-excited carriers in low-temperature ordered states with band gap opening and formulate carrier relaxation time $\tau$ near and below transition temperature $T_{\mathrm{c}}$ by quantifying contributions from different carrier-phonon scatterings to the relaxation rate. The theory explains anomalous experimental observations of $\tau$ in gapped systems. Transverse acoustic (TA) phonon modes play a crucial role in carrier relaxation; their heat capacity determines $\tau$-divergence near $T_{\mathrm{c}}$. The theory is validated by fitting $\tau$ of fullerene polymers onto a theoretical curve.
\end{abstract}


\maketitle

The nonequilibrium dynamics of photo-excited carriers in solids has attracted considerable research interest in the field of condensed matter physics. These dynamics are governed by multiple scatterings of carriers and energy transfer to the phonon field, both of which are quantified by the carrier relaxation time $\tau_{\mathrm{exp}}$ and its temperature ($T$) dependence. Usually anomalous $T$-dependence of $\tau_{\mathrm{exp}}$ is observed in many {\it gapped systems}, for example, phase-ordered systems showing an energy gap opening in the electron band below the transition temperature $T_{\mathrm{c}}$. It has been found that a wide variety of superconductors \cite{kabanov, smith, fleischer, demsar1, kabanov2, chia3, naito, liu, mertelj, chia, coslovich, toda3} and density-wave compounds \cite{demsar2, demsar3, chia2, shimatake, yusupov, toda, toda2} show diverging behavior of $\tau_{\mathrm{exp}}$ near $T_\mathrm{c}$, as confirmed by femtosecond time-resolved optical spectroscopy \cite{kabanov, demsar1, kabanov2, naito, liu, mertelj, chia, coslovich, toda3, demsar2, demsar3, chia2, shimatake, yusupov, toda2}. The divergence of $\tau_{\mathrm{exp}}$ in these gapped systems is believed to result from recursive energy transfer between electrons and phonons. Photo-excited electrons having high energy emit a number of phonons through relaxation from above to below the energy gap. Conversely, relaxed electrons can be re-excited above the gap by absorbing phonon energy. This phonon emission-reabsorption process becomes efficient near $T_{\mathrm{c}}$ because of the small gap energy, and thus, it suppresses the relaxation of carriers, extending $\tau_{\mathrm{exp}}$ significantly. This anomalous phonon-mediated relaxation in gapped systems is called the phonon bottleneck effect. This bottleneck enables the reproduction of these experimental observations of $\tau_{\mathrm{exp}}$ in various gapped systems that exhibit $\tau_{\mathrm{exp}}$-divergence at $T_\mathrm{c}$.

There exists, however, a distinct class of gapped systems such as Tl-based superconductors \cite{smith,chia3} and C$_{60}$-related materials \cite{fleischer, toda} that, instead of showing $\tau_{\mathrm{exp}}$-divergence at $T_{\mathrm{c}}$, show monotonic increases in $\tau_{\mathrm{exp}}$ with cooling across $T_{\mathrm{c}}$. In light of the bottleneck, the monotonic variation in $\tau_{\mathrm{exp}}$ near $T_{\mathrm{c}}$ appears controversial, leading to the questions of why the $\tau_{\mathrm{exp}}$-divergence vanishes in a portion of gapped systems despite well-defined energy gap formation at the Fermi level and whether the bottleneck concept is completely invalid in these gapped systems. Theoretical studies have been unable to satisfactorily answer these two questions over the last decade.

In this Letter, we develop a theory of photo-excited carrier relaxation dynamics with the objective of resolving the abovementioned issues. We postulate that the lack of $\tau_{\mathrm{exp}}$-divergence even in gapped systems can be ascribed to the presence of transverse acoustic (TA) phonon modes. Further, we state that an ensemble of TA modes in gapped systems serves as a high-capacity thermal sink, and energy release to the sink from other phonon modes facilitate efficient cooling of carriers, in other words, a significant reduction in $\tau_{\mathrm{exp}}$ close to $T_{\mathrm{c}}$. The proposed theory is validated by the quantitative agreement with experimental data of $\tau_{\mathrm{exp}}$ for peanut-shaped C$_{60}$ polymers, a typical charge density wave (CDW) material that does not show any carrier relaxation anomalies.

\begin{figure}
\center
\includegraphics[scale=0.33,clip]{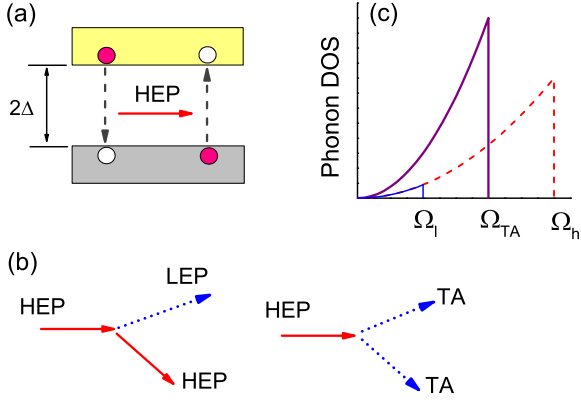}
\caption{\label{fig:density}(Color online) (a) Emission-reabsorption process of HEPs across the energy gap with width $2\Delta$. (b) Diagram of three-phonon scattering processes through which HEPs dissipate. Only the scatterings that involve TA modes are relevant to the lack of divergence in $\tau$. (c) Phonon density of states of TA modes with cutoff frequency $\Omega_{\mathrm{TA}}$, and those of LA modes with $\Omega_h$. The maximum frequency of the LEP is $\Omega_l$ (see text for its definition).}
\end{figure}

First, we briefly review the conventional bottleneck concept for $\tau_{\mathrm{exp}}$-divergence. It is based on the assumption that photo-excited carrier relaxation in gapped systems is regulated by the anharmonic slow decay of phonons \cite{kabanov,rothwarf}. By the absorption of a pump laser photon, electrons in the valence band are excited far above the initial states, and they rapidly accumulate in the upper end of the gap through carrier-carrier and carrier-phonon interactions. Then, the carriers emit high-energy phonons (HEPs) whose energies are higher than the gap width $2\Delta$ to relax into the lower end of the gap. The HEPs produced can be reabsorbed to create new carriers above the gap [see Fig.~\ref{fig:density}(a)] or they can decay in an anharmonic manner into low-energy phonons (LEPs) that no longer excite carriers because their energy is lower than $2\Delta$ [see left-hand side panel in Fig.~\ref{fig:density}(b)]. If the reabsorption probability per unit time is much larger than the inverse of the anharmonic decay time ($\tau^{-1}$), photo-excited carriers and HEPs settle in nearly steady states that obey the Fermi and Bose distribution functions, respectively, with temperature $T'$ that is higher than the lattice temperature $T$ \cite{kabanov}. Meanwhile, LEPs obey the Bose distribution function with $T$, because they remain unperturbed after the laser pulse incident. Consequently, the carrier relaxation is dominated by the energy transfer from HEPs to LEPs, that is, $\tau_{\mathrm{exp}} \simeq \tau$, which is described by the time evolution of the two temperatures, that is, $T'=T'(t)$ and $T=T(t)$.

It should be emphasized that in the conventional bottleneck concept, only longitudinal acoustic (LA) phonon modes are considered. Here, we point out the unnoticed but important role of TA phonon modes in the HEP's decay [see right-hand side panel in Fig.~\ref{fig:density}(b)]. Because TA phonon modes generate no density modulation in the lattice, they cannot interact with photo-excited carriers within the deformation potential theory. This fact implies that TA modes serve as a thermal receiver into which HEPs can dissipate, as a result of which the $\tau$-divergence vanishes even at $T_{\mathrm{c}}$. Below, we prove that this holds true in certain gapped systems.

To formulate the anharmonic decay time $\tau$ of HEPs, we consider the time evolution of the phonon distribution function. Following the above discussion, we divide phonon excitations into two groups. One group consists of HEPs ($\Omega_{l} < \omega_{\bm{q},\mathrm{LA}}< \Omega_{h}$) in equilibrium temperature at $T'$ that can participate in the reabsorption process. The other group involves TA modes ($\omega_{\bm{q},\mathrm{TA}} < \Omega_{\mathrm{TA}}$) and LEPs [$\omega_{\bm{q},\mathrm{LA}} < 2\Delta/\hbar (\equiv\Omega_{l})$] in equilibrium at $T$ that do not participate in the reabsorption process. Here, $\omega_{\bm{q},\mathrm{TA}}$ and $\omega_{\bm{q},\mathrm{LA}}$ represent the phonon dispersion relations, where $\bm{q}$ is the wavevector. $\Omega_{h}, \Omega_{l},$ and $\Omega_{\mathrm{TA}}$ are the cutoff frequencies for the HEP, LEP, and TA modes within the Debye approximation, respectively. The distribution functions for the two groups are given by
\begin{eqnarray}
 n(\omega_{\bm{q},j}) =
 \left[ \exp{\left(\frac{\hbar\omega_{\bm{q},j}}{k_B x}\right)}-1 \right]^{-1},
 \label{eq:noneq_distribution}
\end{eqnarray}
with an appropriate variable $x=T$ or $T'$.

The rate of change in $n(\omega_{\bm{q},j})$ due to phonon-phonon collisions is written as \cite{ziman}
\begin{eqnarray}
 \frac{\partial n(\omega_{\bm{q},j})}{\partial t}
 = J_{\mathrm{col}}[n(\omega_{\bm{q},j})].
\end{eqnarray}
The collision integral $J_{\mathrm{col}}[n(\omega_{\bm{q},j})]$ describes three-phonon scattering, and it is defined by
\begin{eqnarray}
 J_{\mathrm{col}}[n(\omega_{0})] = \frac{2\pi}{\hbar N^2}
   \sum_{\bm{q}_1, \bm{q}_2, j_1, j_2}
   \left\vert w_{i\rightarrow f} \right\vert^2
  \left( \frac{1}{2} S_{A} + S_{B} \right),
\end{eqnarray}
where
\begin{widetext}
\begin{eqnarray}
 S_{A} &=& \left\{ [n(\omega_0)+1] n(\omega_1) n(\omega_2)
 - n(\omega_0) [n(\omega_1)+1][n(\omega_2)+1] \right\}
  \delta(\hbar \omega_0 -\hbar \omega_1-\hbar \omega_2),\\
 S_{B} &=& \left\{ [n(\omega_0)+1][n(\omega_1)+1]n(\omega_2)
 - n(\omega_0) n(\omega_1)[n(\omega_2)+1] \right\}
  \delta(\hbar \omega_2 -\hbar \omega_0 - \hbar \omega_1).
 \label{eq:sb}
\end{eqnarray}
\end{widetext}
Here, $\omega_s (s=0,1,2)$ is an abbreviation of $\omega_{\bm{q}_s,j_s}$, and $w_{i\rightarrow f}$ is the matrix element between the initial and the final state \cite{scatt}.

Note that $J_{\mathrm{col}}$ regulates the time evolution of the total energy through the relationship
\begin{eqnarray}
 \frac{\partial}{\partial t}(E_{\mathrm{LA}}+E_{\mathrm{TA}})
= \sum_{\bm{q} (\mathrm{LA})} \hbar\omega_{\bm{q},\mathrm{LA}} J_{\mathrm{col}}
+ \sum_{\bm{q} (\mathrm{TA})} \hbar\omega_{\bm{q},\mathrm{TA}}
  J_{\mathrm{col}},
 \nonumber\\
 \label{eq:energy_rate_equation}
\end{eqnarray}
where the first and second summations on the right-hand side run over $\bm{q}$s satisfying $0<\omega_{\bm{q},\mathrm{LA}}<\Omega_l$ and $0<\omega_{\bm{q},\mathrm{TA}}<\Omega_{\mathrm{TA}}$, respectively. The energy $E_{j}$ is defined by
\begin{eqnarray}
 E_{j}
= \sum_{\bm{q} (j)} \hbar\omega_{\bm{q},j} n(\omega_{\bm{q},j})
= \int_{0}^{y} \hbar\omega n(\omega)\rho_{j} (\omega) d\omega,
 \label{eq:low_energy}
\end{eqnarray}
where $y=\Omega_{\mathrm{TA}}(\Omega_{l})$ for $j=$TA (LA). $\rho_{j}$ is the density of states defined by
\begin{eqnarray}
 \rho_{j} (\omega) = \alpha_{j} \omega^{2}
 \theta (z - \omega), \
 \left( \alpha_{j} = \frac{3\nu_j N}{z^{3} \sum_{j'}\nu_{j'}} \right),
 \label{eq:dos}
\end{eqnarray}
where $z=\Omega_{\mathrm{TA}}(\Omega_h)$ for $j=$TA (LA), $\nu_{j'}$ is the number of phonon branches participating in the three-phonon scattering, $N$ is the number of unit cells, and $\theta$ is the Heaviside step function; the total density of states is given by $\rho = \rho_{\mathrm{TA}}+\rho_{\mathrm{LA}}$ [see Fig.~\ref{fig:density}(c)], satisfying the normalization condition $1=\int \rho (\omega) d\omega/N$. The Debye approximation we have used in Eq.~(\ref{eq:dos}) is valid when $k_BT<\hbar \Omega_{\mathrm{TA}}$($<\hbar \Omega_{h}$), which holds for many gapped systems below $T_{\mathrm{c}}$. We simplify $n(\omega)$ in Eq.~(\ref{eq:low_energy}) as $k_BT/\hbar\omega$ \cite{kabanov} to obtain $E_{j}=C_{j}T$, where $C_{j}=k_B \alpha_{j} y^{3}/3$ ($k_B$ is the Boltzmann constant). In addition, we focus on the fact that each sum in Eq.~(\ref{eq:energy_rate_equation}) gives a non-zero value only when one or two HEPs contribute to the three-phonon scattering represented by $J_{\mathrm{col}}$ [see Fig.~\ref{fig:density}(b) as an example]. As a result, Eq.~(\ref{eq:energy_rate_equation}) is rewritten as \cite{note}
\begin{eqnarray}
 \frac{\partial T(t)}{\partial t} &=& \frac{1}{\tau}[T' - T(t)],
 \label{eq:temp_rate_equation} \\
 \tau &=& \frac{C_{\mathrm{LA}}+C_{\mathrm{TA}}}
               {I_{\mathrm{LA}} + I_{\mathrm{TA}}},
 \label{eq:tau_definition}
\end{eqnarray}
using the definitions
\begin{eqnarray}
 I_{\mathrm{LA}} &=& w_{1}^{2} \gamma_0 \left( V_A T + V_B T' \right),
 \label{eq:collision1} \\
 I_{\mathrm{TA}} &=& w_{2}^{2} \gamma_1 \left( V_C T + V_D T' \right)
        + w_{2}^{2} \gamma_2 V_E T,
 \label{eq:collision2}
\end{eqnarray}
where $\gamma_k = 2\pi k_{B}^{2} \alpha_{\mathrm{LA}}^{3-k} \alpha_{\mathrm{TA}}^{k}/(\hbar^3 N^2)$. $T'$ is given by
\begin{eqnarray}
 k_BT' = -\frac{\Delta}
 {\ln \left( \epsilon + e^{-\Delta/k_BT} \right)},
\end{eqnarray}
where $\epsilon$ is the dimensionless photoexcitation energy \cite{kabanov}. To derive Eqs.~(\ref{eq:collision1}) and (\ref{eq:collision2}), we assumed that the matrix element is momentum independent, that is, $w_{i\rightarrow f}\equiv w_{1}$ (or $w_{2}$) $=$const. when TA modes are absent from (join in) the three-phonon scattering. $V_{X}$ ($X=A$ to $E$) in Eqs.~(\ref{eq:collision1}) and (\ref{eq:collision2}) are functions of $\Omega_l, \Omega_{\mathrm{TA}}$, and $\Omega_{h}$. 

\begin{figure}[tt]
\center
\includegraphics[scale=0.3,clip]{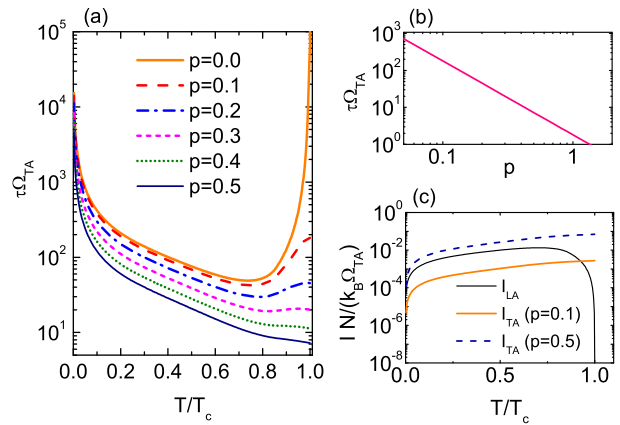}
\caption{\label{fig:tau}(Color online) (a) Numerical result of $\tau (T)$ based on Eq. (\ref{eq:tau_definition}). See text for detailed numerical conditions. (b) Inverse square law of $\tau$ at $T=T_\mathrm{c}$ with respect to $p$. (c) $T$-dependence of $I_{\mathrm{LA}}$ and $I_{\mathrm{TA}}$ defined by Eqs.~(\ref{eq:collision1}) and (\ref{eq:collision2}). Because $I_{\mathrm{TA}}$ differs from zero at $T=T_\mathrm{c}$, the $\tau$-divergence at $T_\mathrm{c}$ is strongly suppressed.}
\end{figure}

Equation (\ref{eq:tau_definition}) is the main finding of this study, because it clearly shows the contribution of TA phonon modes to carrier relaxation. Figure~\ref{fig:tau}(a) shows the $T$-dependence of $\tau$ that we have formulated in Eq.~(\ref{eq:tau_definition}). The parameter $p\ (\equiv w_2/w_1)$ is the coupling strength through which the HEPs decay into TA modes, and it is tuned from 0 to 0.5 in increments of 0.1. With an increase in $p$, the magnitude of $\tau$ decreases over the entire $T$ range. When $p$ is smaller (larger) than $\sim 0.3$, $\tau$ increases (decreases) as $T$ approaches $T_{\mathrm{c}}$ from below. The most significant phenomenon is the drastic reduction in $\tau$ at $T_{\mathrm{c}}$ with increasing $p$. In fact, $\tau$ at $T_{\mathrm{c}}$ is inversely proportional to $p^2$, as shown in Fig.~\ref{fig:tau}(b). These results indicate that the lack of $\tau$-divergence at $T_c$ is attributable to efficient phonon-phonon coupling between the HEP and the TA mode characterized by $p$.

To explain the microscopic mechanism for the lack of divergence, in Fig.~\ref{fig:tau}(c), we show the $T$-dependence of the magnitude of $I_{\mathrm{LA}}$ and $I_{\mathrm{TA}}$ given in Eqs.~(\ref{eq:collision1}) and (\ref{eq:collision2}). Here, $I_{\mathrm{LA(TA)}}$ quantifies the efficiency of the HEP's energy dissipation through the anharmonic interaction with LA (TA) phonon modes. In the limit of $T\rightarrow T_{\mathrm{c}}$ (that is, $\Delta\rightarrow 0$), $I_{\mathrm{LA}}$ vanishes (irrespective of $p$) but $I_{\mathrm{TA}}$ converges to a finite value as long as $p\ne 0$. Therefore, we obtain a non-diverging $\tau$ at $T_{\mathrm{c}}$ if $p\ne 0$, which readily follows from Eq.~(\ref{eq:tau_definition}). When $p=0$, on the other hand, $I_{\mathrm{TA}}\equiv 0$ for arbitrary $T$ [because $I_{\mathrm{TA}} \propto w_{2}^{2}$; see Eq.~(\ref{eq:collision2})]. In this case, $I_{\mathrm{LA}} = I_{\mathrm{TA}}= 0$ at $T_{\mathrm{c}}$ so that $\tau \rightarrow \infty$ at $T_{\mathrm{c}}$. We thus conclude that the anharmonic decay of HEPs to TA modes plays a prominent role in the efficient cooling of HEPs as well as in determining the significance of the $\tau$-divergence at $T_{\mathrm{c}}$.

Upon reducing the temperature to zero, the energy dissipation of HEPs gradually reduces due to monotonic decreases in $I_{\mathrm{LA}}$ and $I_{\mathrm{TA}}$ [see Fig.~\ref{fig:tau}(c)]. The decrease in $I_{\mathrm{LA(TA)}}$ is attributed to the reduced phonon population, which results in a monotonic increase in $\tau$ at low $T$, as shown in Fig.~\ref{fig:tau}(a). A similar increase has been found in various gapped systems \cite{smith, fleischer, demsar1, kabanov2, chia, coslovich, toda3, chia2, toda}, and it is attributable to the inefficient cooling of HEPs.

\begin{figure}[tt]
\center
\includegraphics[scale=0.3,clip]{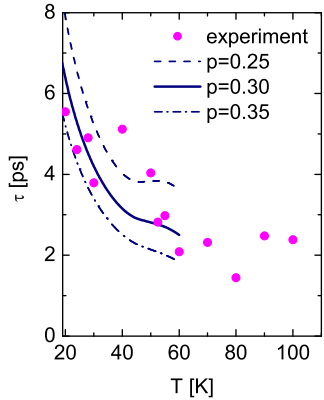}
\caption{\label{fig:tau2}(Color online) Experimental data of $\tau_{\mathrm{exp}}$ in the C$_{60}$ polymer (after Ref.~\cite{toda}) and their numerical reproduction based on Eq. (\ref{eq:tau_definition}). The data for $T\le T_{\mathrm{c}}=60$K fits the theoretical curve with $p=0.3$.}
\end{figure}

Now, we apply the proposed theory to photo-excited carrier relaxation in quasi one-dimensional C$_{60}$ polymers. It was previously observed in experiments \cite{onoe} that the C$_{60}$ polymers undergo the CDW transition \cite{ono2} at $T_{\mathrm{c}}=60$K, forming a well-defined energy gap at the Fermi level. This result implied the possibility of $\tau_{\mathrm{exp}}$-divergence at $60$K; nevertheless, optical pump-probe investigations \cite{toda} of the C$_{60}$ polymers revealed a monotonic variation in $\tau_{\mathrm{exp}}$ near $T_{\mathrm{c}}$, whose origin has yet to be clarified. This problem is solved by considering that the twisting phonon modes of the C$_{60}$ polymer \cite{ono1} play the same role as the above-described TA modes. Figure~\ref{fig:tau2} shows the numerical reproduction (indicated by lines) of the experimental data \cite{toda} (indicated by circles) of the $T$-dependent $\tau_{\mathrm{exp}}$ of the C$_{60}$ polymers. An overall agreement between the theory and the experiments is obtained by assuming $p\sim 0.3$. The other parameters we used are $\Omega_{\mathrm{TA}}=220$cm$^{-1}$, $\Omega_{h}=360$cm$^{-1}$, and $2\Delta(0)=360$K; the first two values were estimated from the phonon model for 1D C$_{60}$ polymers \cite{ono1}, and the last value gives $2\Delta(0)/k_BT_c=6$ consistent with many CDW compounds \cite{gruner}.

The generality of Eq. (\ref{eq:tau_definition}) is of great importance. It is applicable to $\tau_{\mathrm{exp}}$ in other gapped systems such as Tl-based cuprate superconductors \cite{smith,chia3} and the solid fullerenes K$_{3}$C$_{60}$ and Rb$_{3}$C$_{60}$ \cite{fleischer}. Even these materials show a lack of $\tau_{\mathrm{exp}}$-divergence; however, no theoretical studies have attempted to clarify their relaxation anomalies. We believe that the proposed theory will serve as a unified framework for nonequilibrium carrier dynamics in gapped systems.

In conclusion, we have developed a theory to describe the energy dissipation from HEPs ($\hbar\omega_{\bm{q},\mathrm{LA}}>2\Delta$) to LEPs ($\hbar\omega_{\bm{q},\mathrm{LA}}<2\Delta$) and TA modes in gapped systems below $T_\mathrm{c}$. This theory enables the evaluation of the $T$-dependence of $\tau$ as a function of the coupling strength $p\equiv w_2/w_1$ between the HEPs and the TA modes, and it explains the crossover between the diverging and the non-diverging behaviors of photo-excited carrier relaxation. The latter behavior can quantitatively account for the anomalous carrier relaxation time in C$_{60}$ polymers, as measured by recent pump-probe laser experiments. The coupling between the HEPs and the TA mode also suggests the variation of $\tau_{\mathrm{exp}}$-divergence in typical gapped systems.

We thank K. Yakubo, Y. Asano, J. Onoe, K. Ohno, Y. Noda, and T. Matsuura for the helpful discussions. This study was supported by a Grant-in-Aid for Scientific Research from the MEXT, Japan. HS acknowledges the financial support from the Inamori Foundation and the Sumitomo Foundation.

\nocite{*}



\begin{thebibliography}{99}




\bibitem{kabanov} V. V. Kabanov, J. Demsar, B. Podobnik, and D. Mihailovic, Phys.\ Rev. B {\bf 59}, 1497 (1999).

\bibitem{smith} D. C. Smith, P. Gay, D. Z. Wang, J. H. Wang, Z. F. Ren, and J. F. Ryan, Physica C (Amsterdam) {\bf 341}, 2219 (2000).

\bibitem{fleischer} S. B. Fleischer, B. Pevzner, D. J. Dougherty, H. J. Zeiger, G. Dresselhaus, M. S. Dresselhaus, E. P. Ippen, and A. F. Hebard, Phys.\ Rev. B {\bf 62}, 1366 (2000).

\bibitem{demsar1} J. Demsar, R. Hudej, J. Karpinski, V. V. Kabanov, and D. Mihailovic, Phys.\ Rev. B {\bf 63}, 054519 (2001).

\bibitem{kabanov2} V. V. Kabanov, J. Demsar, and D. Mihailovic, Phys. Rev. Lett. {\bf 95}, 147002 (2005).

\bibitem{chia3} E. E. M. Chia, J.-X. Zhu, D. Talbayev, R. D. Averitt, A. J. Taylor, K.-H. Oh, I.-S. Jo, and S.-I. Lee, Phys. Rev. Lett. {\bf 99}, 147008 (2007).

\bibitem{naito} T. Naito, Y. Yamada, T. Inabe, and Y. Toda, J. Phys. Soc. Jpn. {\bf 77}, 064709 (2008).

\bibitem{liu} Y. H. Liu, Y. Toda, K. Shimatake, N. Momono, M. Oda, and M. Ido, Phys. Rev. Lett. {\bf 101}, 137003 (2008).

\bibitem{mertelj} T. Mertelj, P. Kusar, V. V. Kabanov, L. Stojchevska, N. D. Zhigadlo, S. Katrych, Z. Bukowski, J. Karpinski, S. Weyeneth, and D. Mihailovic, Phys.\ Rev. B {\bf 81}, 224504 (2010).

\bibitem{chia} E. E. M. Chia, D. Talbayev, J.-X. Zhu, H. Q. Yuan, T. Park, J. D. Thompson, C. Panagopoulos, G. F. Chen, J. L. Luo, N. L. Wang, and A. J. Taylor, Phys. Rev. Lett. {\bf 104}, 027003 (2010).

\bibitem{coslovich} G. Coslovich, C. Giannetti, F. Cilento, S. Dal Conte, G. Ferrini, P. Galinetto, M. Greven, H. Eisaki, M. Raichle, R. Liang, A. Damascelli, and F. Parmigiani, Phys.\ Rev. B {\bf 83}, 064519 (2011).

\bibitem{toda3} Y. Toda, T. Mertelj, P. Kusar, T. Kurosawa, M. Oda, M. Ido, and D. Mihailovic, Phys.\ Rev. B {\bf 84}, 174516 (2011).



\bibitem{demsar2} J. Demsar, K. Biljakovi\'{c}, and D. Mihailovic, Phys. Rev. Lett. {\bf 83}, 800 (1999).

\bibitem{demsar3} J. Demsar, L. Forr\'{o}, H. Berger, and D. Mihailovic, Phys.\ Rev. B {\bf 66}, 041101 (2002).

\bibitem{chia2} E. E. M. Chia, J.-X Zhu, H. J. Lee, N. Hur, N. O. Moreno, E. D. Bauer, T. Durakiewicz, R. D. Averitt, J. L. Sarrao, and A. J. Taylor, Phys.\ Rev. B {\bf 74}, 140409(R) (2006).

\bibitem{shimatake} K. Shimatake, Y. Toda, and S. Tanda, Phys.\ Rev. B {\bf 73}, 153403 (2006); Phys.\ Rev. B {\bf 75}, 115120 (2007).

\bibitem{yusupov} R. V. Yusupov, T. Mertelj, J.-H. Chu, I. R. Fisher, and D. Mihailovic, Phys. Rev. Lett. {\bf 101}, 246402 (2008).

\bibitem{toda} Y. Toda, S. Ryuzaki, and J. Onoe, Appl.\ Phys.\ Lett. {\bf 92}, 094102 (2008).

\bibitem{toda2} Y. Toda, R. Onozaki, M. Tsubota, K. Inagaki, and S. Tanda, Phys.\ Rev. B {\bf 80}, 121103 (2009).

\bibitem{rothwarf} A. Rothwarf and B. N. Taylor, Phys. Rev. Lett. {\bf 19}, 27 (1967).

\bibitem{ziman} J. M. Ziman, {\it Electrons and Phonons} (Oxford University Press, New York, 2001).

\bibitem{scatt} For example, the first term in the curly brackets in Eq.~(\ref{eq:sb}) implies that one phonon with $\omega_2$ decays into two phonons with $\omega_0$ and $\omega_1$.

\bibitem{note} In the calculation, we assume that $T'$ is fixed [that is, $T'(0) \simeq T'(\infty)$] in order to derive the analytical expression of $\tau$. A similar assumption has been used in Ref.~\cite{kabanov} for the case of the absence of the TA mode.

\bibitem{onoe} J. Onoe, A. Takashima, and Y. Toda, Appl.\ Phys.\ Lett. {\bf 97}, 241911 (2010).

\bibitem{ono2} S. Ono and H. Shima, EPL {\bf 96}, 27011 (2011).

\bibitem{ono1} S. Ono and H. Shima, J. Phys. Soc. Jpn. {\bf 80}, 064704 (2011).

\bibitem{gruner} G. Gr\"{u}ner, {\it Density Waves in Solids} (Addison-Wesley, MA, 1994).









\end{thebibliography}
\end{document}